\UseRawInputEncoding
 \documentclass[%
reprint,
superscriptaddress,
showpacs,
preprintnumbers,
amsmath,amssymb,
aps,
prl,
]{revtex4-1}

\usepackage{graphicx}
\usepackage{dcolumn}
\usepackage{bm}
\usepackage{xcolor}
\usepackage[colorinlistoftodos]{todonotes}
\usepackage{float}

\begin{document}

\newpage

\title{Phase-controlled lateral forcing of droplets on a vibrating bath}
\author{Lawrence A. Wilen}
\author{Zahra Shamsi}
\thanks{These authors contributed equally to this work.}
\author{Johnathan Hoggarth}
\thanks{These authors contributed equally to this work.}
\author{Bauyrzhan K. Primkulov}
\email{bauyrzhan.primkulov@yale.edu}

\affiliation{Mechanical Engineering, Yale University}

\date{\today}

\begin{abstract}
We introduce a method for controlling the lateral motion of droplets bouncing on a vibrated liquid bath by superimposing a small periodic rocking motion onto the conventional vertical forcing. We show experimentally that the lateral velocity varies linearly with the rocking amplitude $\alpha$ and sinusoidally with its phase $\phi$ relative to the vertical forcing, while the response depends strongly on the droplet's vertical bouncing mode. A time-averaged model of a harmonic bouncer yields a time-averaged lateral force $\overline{F}_x=mg\alpha\sin(\Phi_b+\phi)$, where $\Phi_b$ is the droplet impact phase. Periodic rocking thus provides a simple, programmable means of exerting lateral forces on bouncing droplets, enabling controlled manipulation of passive and active droplets in pilot-wave hydrodynamics.
\end{abstract}

\maketitle

\textbf{\label{sec:Introduction}Introduction.}
Pilot-wave hydrodynamics features a unique class of active particles whose activity and interactions arise from a self-excited wave field~\citep{couder_bouncing_2005,bush_pilot-wave_2015,bush_hydrodynamic_2021}. A droplet bouncing on the surface of a vertically vibrated liquid bath generates periodic disturbances at the free surface. As the peak driving acceleration $\gamma$ approaches the Faraday instability threshold $\gamma_F$, these disturbances excite slowly decaying subharmonic waves. When the droplet itself bounces subharmonically, it can transition from a passive bouncing state to an active walking state, self-propelling across the bath together with its guiding wave field~\citep{couder_bouncing_2005, protiere_particlewave_2006, molacek_drops_2013b, oza_trajectory_2013}.

The resulting particle–wave entity exhibits a remarkable range of behaviors reminiscent of quantum phenomena~\citep{bush_hydrodynamic_2021}, including quantized orbital states~\citep{fort_path-memory_2010, perrard_self-organization_2014, harris_droplets_2014, oza_pilot-wave_2014} and wave-like statistics in confined geometries~\citep{harris_wavelike_2013,saenz_statistical_2017}. Importantly, some of these landmark experiments have relied on externally imposed forces to control the horizontal motion of the droplet. Quantized orbital states, for example, have been realized by subjecting droplets to a Coriolis force~\citep{fort_path-memory_2010} or by embedding a ferrofluid core within the droplet and using a magnetic field to generate a central confining potential~\citep{perrard_self-organization_2014}. These experiments illustrate how controlled lateral forcing can provide a powerful means of manipulating pilot-wave dynamics.

A more ubiquitous source of lateral forcing has traditionally been regarded as an experimental imperfection. Slight misleveling of the bath or unintended vibrational modes that depart from purely vertical forcing can produce persistent drift in droplet trajectories~\citep{bush_pilot-wave_2015}. Considerable experimental care is therefore typically taken to level the bath and ensure uniform vertical forcing~\citep{harris_generating_2015}. Here, we instead exploit this effect by deliberately superimposing a small periodic rocking motion onto the conventional vertical forcing. The resulting bath inclination generates a lateral component of the drop--bath impact force whose magnitude and direction are controlled by the timing of impact relative to the rocking cycle. By independently varying the amplitude and phase of the rocking motion, we demonstrate tunable propulsion of bouncing droplets in prescribed horizontal directions. Thus, a vibrational mode normally suppressed as an experimental imperfection becomes a programmable means of exerting lateral forces on bouncing droplets.

\begin{figure}
    \centering
    \includegraphics[width=\linewidth]{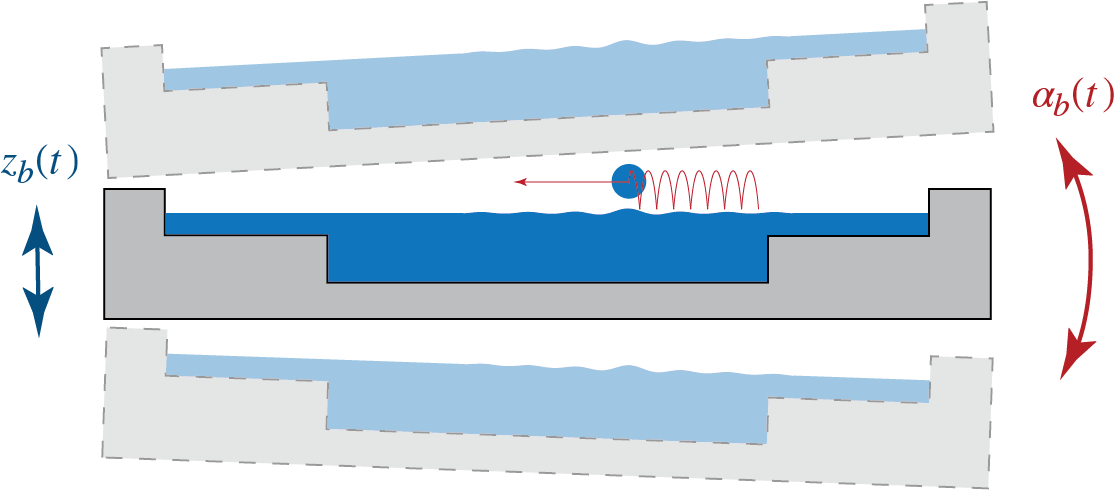}
    \caption{Schematic of the vibrationally forced silicone-oil bath. The vertical displacement at the center of the bath is prescribed as $z_b(t)=-(\gamma/\omega^2)\sin(\omega t)$, where $\omega=2\pi f$. In addition to this vertical motion, the bath undergoes a very small periodic tilt at the same frequency, with inclination angle $\alpha_b(t)=-\alpha\sin(\omega t+\phi)$, where $\alpha$ is the tilt amplitude and $\phi$ is its phase relative to the vertical forcing. The oscillatory tilt imposes a lateral impulse on the droplet at each impact, enabling controlled propulsion in a prescribed direction.}
    \label{fig:setup}
\end{figure}

\textbf{Experimental Setup.}
The experimental setup consists of four independently driven speakers arranged around a circular platform supporting a square bouncing-droplet cell ({37 mm wide}). By controlling the relative amplitudes and phases of the four speakers at a common driving frequency of $f=80$\;Hz, the platform motion can be decomposed into uniform vertical oscillation and two orthogonal rocking modes, allowing independent control of the platform’s linear acceleration and tilt (see Fig.~\ref{fig:setup} and \citep{noauthor_see_nodate} for additional details). Differences in speaker response are removed through calibration using a six-axis inertial measurement unit mounted beneath the cell, with accelerometers measuring linear motion and gyroscopes measuring rocking motion; the sensor signals are analyzed using a software lock-in technique and calibrated against direct displacement measurements obtained with high-speed imaging. 

At the experimental rocking frequency of $f=80$ Hz, which is significantly higher than the natural frequency of free-surface sloshing within the cell, the slope of the liquid surface closely follows that of the cell (see \citep{noauthor_see_nodate} for details). The addition of rocking has two effects: (i) it periodically varies the overall bath slope, and (ii) it introduces a spatial variation in the peak vertical acceleration, which is lower near speakers driven at lower amplitude and higher near those driven at higher amplitude (see \citep{noauthor_see_nodate}). We minimize the latter effect by using a relatively small bath and restricting measurements to its central region.

Experiments typically begin with pure vertical forcing, after which a droplet of prescribed radius ($R=0.32 \pm 0.02$ mm or $R=0.43 \pm 0.01$ mm) is generated using a click-pen mechanism~\citep{hoggarth_controlled_2026} and introduced onto the bath. Controlled rocking is then used to propel the droplet in any desired horizontal direction. In a typical measurement (\textcolor{red}{Video 1}), the droplet is driven repeatedly across the cell in opposite directions, with its velocity measured as it crosses the center, where the peak vertical acceleration is $\gamma$, the rocking produces no local linear acceleration, and wall effects are minimized. All experiments were performed using a $20$\;cSt silicone-oil bath (density $\rho=950$\;kg/m$^3$, surface tension $\sigma=20.1$\;mN/m), with a depth of $8.5$ mm and a surrounding shallow region $1$ mm deep. Complete details of the experimental setup are provided in the supplemental material~\citep{noauthor_see_nodate}.

\textbf{Experimental observations.}
While the droplet can be propelled in any direction along the bath surface (\textcolor{red}{Video 2}), without loss of generality, we restrict our attention to motion along one of the bath's horizontal axes of symmetry. We impose a periodic tilt, 
\begin{equation}
    \alpha_b(t)=-\alpha\sin(\omega t+\phi),
    \label{eqn:tilt}
\end{equation}
while the vertical displacement of the bath center follows $z_b(t)=-(\gamma/\omega^2)\sin(\omega t)$.

\begin{figure}
    \centering
    \includegraphics[width=\linewidth]{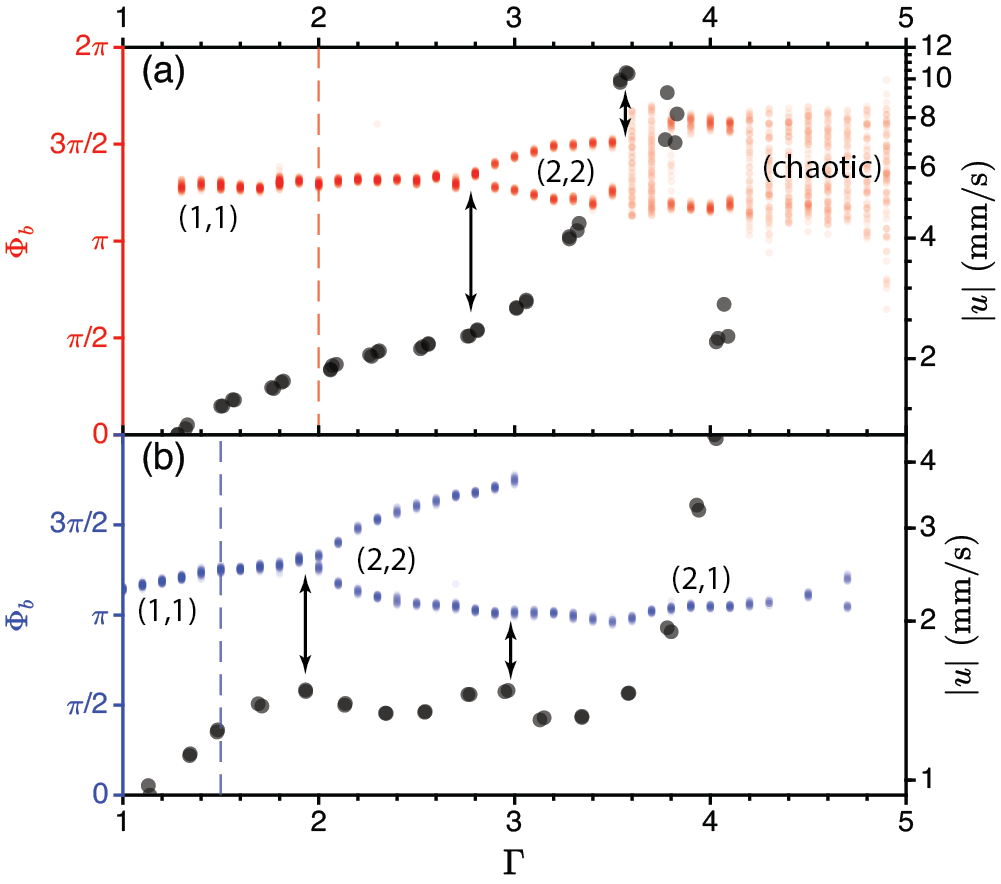}
    \caption{Drop impact phase $\Phi_b$ and speed $|u|$ as a function of the dimensionless driving acceleration $\Gamma$ for (a)~$R=0.43$\;mm and (b)~$R=0.32$\;mm drops. For both (a) and (b) the tilt amplitude was fixed at $\alpha=0.06^\circ$ and phase shift $\phi=17^\circ$. Abrupt changes in the drop speed (shown on a logarithmic scale) coincide with transitions between $(m,n)$ bouncing modes, in which the drop undergoes $n$ bounces over $m$ bath periods. The small arrows indicate the locations of bouncing mode transitions.}
    \label{fig:acceleration_and_impact_phase}
\end{figure}

We first measure when, within each driving period, the drop impacts the bath with no tilt imposed (see \citep{noauthor_see_nodate}). If an impact occurs at time $t_i$, we define the corresponding impact phase as $\Phi_b=\omega t_i$ (mod $2\pi$), such that $\Phi_b=0$ and $\pi$ correspond to impacts when the bath is moving downward and upward at its maximum speed, respectively. Figure~\ref{fig:acceleration_and_impact_phase} shows the measured impact phase as a function of the dimensionless driving acceleration $\Gamma\equiv\gamma/g$~\citep{molacek_drops_2013a, couchman_bouncing_2019}, with the bath rocking disabled. The impact phases were obtained by tracking the vertical motion of the drop and bath using high-speed imaging and are shown for $R=0.43$\;mm (red) and $R=0.32$\;mm (blue).

At low driving accelerations, both drops bounce harmonically, with one impact per forcing period. As $\Gamma$ is increased, both undergo period-doubling transitions, although the transition thresholds depend on drop size~\citep{protiere_particlewave_2006, eddi_wave_2008, molacek_drops_2013a, molacek_drops_2013b, wind-willassen_exotic_2013}. For $R=0.32$\;mm, period doubling occurs at $\Gamma\approx2$, whereas for $R=0.43$\;mm, it occurs at $\Gamma\approx2.8$. At still higher forcing, the dynamics of the two drops diverge: the larger drop eventually exhibits chaotic bouncing~\citep{gilet_chaotic_2009} as the Faraday instability threshold, $\Gamma_F\approx5$, is approached, whereas the smaller drop maintains period-doubled bouncing, with one impact every two forcing periods. As we show below, these distinct bouncing modes strongly influence the lateral motion induced by rocking.

We next fix the tilt amplitude at $\alpha=0.06^\circ$ and the phase lead of the rocking motion at $\phi=17^\circ$ (see Eq.~\eqref{eqn:tilt} and \citep{noauthor_see_nodate}) and measure the drop speed $|u|$ as $\Gamma$ is systematically increased (black dots in Fig.~\ref{fig:acceleration_and_impact_phase}). For the larger drop, $|u|$ initially increases gradually with $\Gamma$, with a more rapid increase following the period-doubling transition. Near $\Gamma\approx3.5$, the drop transitions to a self-propelled state. With further increases in $\Gamma$, chaotic bouncing emerges near $\Gamma\approx4.1$ and is accompanied by a sharp decrease in drop's speed. Remarkably, even in this chaotic regime, periodic rocking continues to propel the drop in a controlled direction. The smaller drop likewise exhibits pronounced changes in speed following transitions between bouncing modes and becomes self-propelled near $\Gamma\approx4$.

For the remainder of the article, we restrict our attention to harmonic bouncing, choosing $\Gamma=2$ for $R=0.43$\;mm and $\Gamma=1.5$ for $R=0.32$\;mm (dashed lines in Fig.~\ref{fig:acceleration_and_impact_phase}). At these driving accelerations, both drops bounce once per forcing period. We then systematically explore how the tilt amplitude $\alpha$ and phase lead $\phi$ in Eq.~\eqref{eqn:tilt} govern their lateral motion.

\begin{figure}
    \centering
    \includegraphics[width=\linewidth]{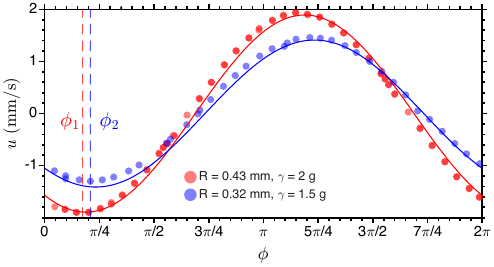}
    \caption{Lateral drop velocity as a function of the phase lead $\phi$ of the rocking motion for a fixed driving acceleration $\gamma$ and tilt amplitude $\alpha=0.06^\circ$. The phase leads $\phi_1$ and $\phi_2$, corresponding to the first two velocity extrema, are indicated by the red and blue dashed lines, respectively. Solid lines correspond to equation~\eqref{eqn:velocity}, using the impact phase $\Phi_b$ measured for each drop size in Fig.~\ref{fig:acceleration_and_impact_phase}.}
    \label{fig:PhaseLead}
\end{figure}

\begin{figure}
    \centering
    \includegraphics[width=\linewidth]{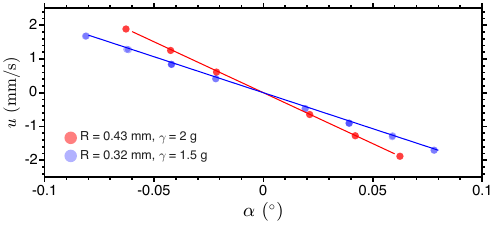}
    \caption{Lateral drop velocity as a function of tilt amplitude $\alpha$ for a fixed driving acceleration $\gamma$. The phase lead is fixed at $\phi=17^\circ$, near the value that maximizes the drop speed. Here, the velocity varies linearly with $\alpha$. Solid lines correspond to equation~\eqref{eqn:velocity}, using the impact phase $\Phi_b$ measured for each drop size in Fig.~\ref{fig:acceleration_and_impact_phase}.}
    \label{fig:RockingMagnitude}
\end{figure}

Fixing the tilt amplitude at $\alpha=0.06^\circ$, we measure the drop velocity $u$ as a function of the phase lead $\phi$ for both drop sizes (Fig.~\ref{fig:PhaseLead}). For both $R=0.32$ mm and $R=0.43$ mm, the velocity varies sinusoidally with $\phi$, although the amplitude and phase of the response depend on drop size. Conversely, fixing the phase lead at $\phi=17^\circ$, we find that the velocity varies linearly with the tilt amplitude $\alpha$ for both drop sizes (Fig.~\ref{fig:RockingMagnitude}). We next introduce a simple model that accounts for these observations, including the sinusoidal dependence on $\phi$, the location of the velocity extrema, and the linear dependence on $\alpha$.

\textbf{Model and Discussion.}
We begin with the non-resonant model of bouncing drops~\citep{molacek_drops_2013a,molacek_drops_2013b,primkulov_nonresonant_2025,damiani_energetics_2026}, in which the drop's vertical motion is governed by
\begin{equation}
m\ddot{z}_p(t) = F_N(t)\cos\alpha_b(t) - mg,
\label{eqn:vertical}
\end{equation}
where $m$ is the drop mass and $F_N(t)$ denotes the normal impact force exerted by the bath on the drop, which is nonzero only while the drop is in contact with the bath. 
The horizontal motion of the drop is governed by
\begin{align}
m\ddot{x}_p(t)
+\left(\mathcal{D}_h\sqrt{\frac{\rho R}{\sigma}}F_N(t)
+6\pi R\mu_a\right)\dot{x}_p(t)
= \nonumber \\
-F_N(t)\left[\partial_x h(x_p,t)+\sin\alpha_b(t)\right],
\label{eqn:horizontal}
\end{align}
where $\mathcal{D}_h\sqrt{\rho R/\sigma} F_N(t)$ represents the horizontal damping during impact, with $\mathcal{D}_h$ denoting the dimensionless damping coefficient, while $6\pi R\mu_a$ accounts for air drag, with $\mu_a$ denoting the air viscosity. In the model of \citet{molacek_drops_2013b}, the horizontal force exerted during impact is proportional to the product of the normal force $F_N$ and the local wave slope $\partial_x h$. We extend this formulation to include the rocking motion of the bath, which introduces an additional overall slope of the bath $\sin\alpha_b(t)$. Consequently, each impact imparts a lateral impulse whose magnitude and direction are determined by the instantaneous bath tilt. The resulting propulsion therefore depends critically on the phase of the drop's impacts relative to the rocking cycle.

The essential physics can be distilled even further by time-averaging equation~\eqref{eqn:horizontal}. We restrict our attention to the periodic harmonic bouncing mode (Fig.~\ref{fig:acceleration_and_impact_phase}), where longevity of waves is comparable to the bouncing period~\citep{molacek_drops_2013b} and we neglect the wave-induced contribution $\partial_x h$ in Eq.~\eqref{eqn:horizontal}. Since the rocking angle is small, $\sin\alpha_b(t)\approx\alpha_b(t)
=-\alpha\sin(\omega t+\phi)$. Furthermore, integrating Eq.~\eqref{eqn:vertical} over one bouncing period $T_b$ and using the periodicity of the vertical motion gives $\frac{1}{T_b}\int_t^{t+T_b}F_N(s) ds=mg$, where we have used $\cos\alpha_b\approx1$. We therefore approximate the impact force by an impulsive force, $F_N(t)\approx mgT_b\delta(t-t_i)$, where $\delta(t-t_i)$ is the Dirac-delta function centered around the impact.

Averaging equation~\eqref{eqn:horizontal} over one bouncing period then gives
\begin{align*}
m\ddot{\bar{x}}_p(t)
+\mathcal{D}_h\sqrt{\frac{\rho R}{\sigma}} mg \dot{x}_p(t_i)
+6\pi R\mu_a\dot{\bar{x}}_p(t) \\
=mg\alpha\sin(\omega t_i+\phi),
\end{align*}
where $\dot{\bar{x}}_p(t) \equiv \frac{1}{T_b} \int_t^{t+T_b} \dot{x}_p(s) ds$ and $\ddot{\bar{x}}_p(t)  \equiv \frac{1}{T_b} \int_t^{t+T_b} \ddot{x}_p(s) ds$ denote the averaged velocity and acceleration, respectively. Approximating the velocity at impact by its period-averaged value, $\dot{x}_p(t_i)\approx\dot{\bar{x}}_p(t)$, and noting that $\omega t_i=\Phi_b$, we obtain
\begin{align}
m\ddot{\bar{x}}_p(t)+b\dot{\bar{x}}_p(t)
=mg\alpha\sin(\Phi_b+\phi),
\label{eqn:averaged}
\end{align}
where $b=\mathcal{D}_h\sqrt{\frac{\rho R}{\sigma}}mg+6\pi R\mu_a$. The steady-state velocity is therefore
\begin{equation}
u=\frac{mg\alpha}{b}\sin(\Phi_b+\phi).
\label{eqn:velocity}
\end{equation}

Here, we take $\Phi_b$ to be experimentally measured average impact phase at a given $\Gamma$ (Fig.~\ref{fig:acceleration_and_impact_phase}). Despite its simplicity, Eq.~\eqref{eqn:velocity} captures the principal experimental trends in Figs.~\ref{fig:PhaseLead}--\ref{fig:RockingMagnitude} when $b$ is treated as a fitting parameter. More importantly, it makes the propulsion mechanism explicit: the period-averaged lateral force, $\overline{F}_x=mg\alpha\sin(\Phi_b+\phi)$, is determined by the phase of the drop impact relative to the rocking motion. The propulsion is therefore predicted to be strongest when the drop impacts at an extremum of the bath tilt, corresponding to $\Phi_b+\phi=\pi/2+\pi k$ (where $k$ is integer), and vanishes when the bath is momentarily level.

It is worth noting that the data and theory do not agree perfectly in their relative phase (see Fig.~\ref{fig:PhaseLead}). This discrepancy likely reflects the simplicity of the impulse approximation, in which the phase of the impact force is identified with the measured time of closest approach between the drop and the bath. Figure~\ref{fig:PhaseLead} provides an alternative perspective: the phase at which the lateral velocity reaches an extremum provides an independent measure of the effective phase of the impact force. This effective phase need not coincide with the impact phase defined by the time of closest approach, as the two quantities are conceptually distinct.

\textbf{Discussion}.
While equation~\eqref{eqn:velocity} captures the dependence of the drop velocity on $\alpha$ and $\phi$, it does not explicitly describe its dependence on $\Gamma$, evident in Fig.~\ref{fig:acceleration_and_impact_phase}. The complexities of the drop--bath interaction, which depend on drop size and impact speed~\citep{galeano-rios_non-wetting_2017, galeano-rios_quasi-normal_2019, gabbard_drop_2025, aguero_droplet_2026}, are instead absorbed into the effective damping coefficient $b$, treated here as a fitting parameter, with fitted values corresponding to $\mathcal{D}_h=0.17$ and $0.11$ for $R=0.32$ mm and $0.43$ mm in eq.~\eqref{eqn:horizontal}, respectively. These values of $\mathcal{D}_h$ are very close to experimentally-estimated values by \citet{molacek_drops_2013b}. Despite this simplification, equation~\eqref{eqn:averaged} provides a simple framework for precisely manipulating passive bouncing drops. Although derived for one-dimensional motion along the $x$-axis, this formulation extends directly to two dimensions, capturing the experimental capability of steering droplets along arbitrary trajectories (\textcolor{red}{Video 3}).

An important feature of this forcing mechanism is that it acts globally: drops of the same mass that bounce in phase experience approximately the same lateral impulse across a sufficiently small bath. This property makes the method particularly well suited to manipulating ensembles of bouncing drops. \citet{eddi_archimedean_2009, eddi_oscillating_2011} showed that bouncing drops can self-organize into crystalline structures on a vibrating bath. The lateral forcing introduced here can therefore be used to translate and deform such crystals (\textcolor{red}{Video 4}).


\textbf{Conclusions}.
We have demonstrated a simple method for controlling the lateral motion of droplets on a vibrated liquid bath by superimposing a small periodic rocking motion onto the conventional vertical forcing. The resulting lateral impulse is highly tunable: the drop velocity varies linearly with the rocking amplitude and sinusoidally with its phase relative to the vertical forcing. Because the direction of the rocking can also be prescribed, this approach enables controlled propulsion of droplets in arbitrary horizontal directions. Moreover, the response depends strongly on the drop's vertical dynamics, with abrupt changes in lateral speed accompanying transitions between harmonic, period-doubled, walking, and chaotic bouncing states.

A time-averaged model captures the principal experimental trends and identifies the physical origin of the lateral forcing. During each impact, the normal force exerted by the bath acquires a horizontal component proportional to the instantaneous bath inclination. The net propulsion is therefore determined by the synchronization between the drop impacts and the rocking motion. For harmonic bouncing, this mechanism reduces to the simple relation [Eq.~\eqref{eqn:velocity}] which directly connects the time-averaged lateral force to the rocking amplitude, rocking phase, and drop impact phase.

Our results thus turn a feature normally treated as an experimental nuisance, departing from purely vertical forcing, into a controllable means of manipulating bouncing droplets. In addition to providing precise control over individual droplet trajectories, this forcing scheme offers a route toward externally manipulating multiple interacting droplets and probing the response of pilot-wave systems to prescribed time-dependent lateral forces.

\textbf{Acknowledgements.} Authors gratefully acknowledge startup funds provided by Yale University and Diana Omar for assistance with early experiments.

\bibliography{references.bib}

\end{document}